\documentclass{ws-p8-50x6-00}

\newcommand{\Mo}{\ensuremath{M_\odot}}

\newcommand{\Lo}{\ensuremath{L_\odot}}

\newcommand{\Lstar}{\ensuremath{L_{\star}}}

\newcommand{\rtLeffrat}{\ensuremath{(L_{\rm eff}/L_{\star})^{1/2}}}

\newcommand{\vstar}{\ensuremath{v_{\star}}}
\newcommand{\vstarsq}{\ensuremath{v_{\star}^{2}}}

\newcommand{\mlb}{\ensuremath{M/L_{B}}}

\newcommand{\kms}{~\ensuremath{{\rm km\ s}^{-1}}}
\newcommand{\hkpc}{~\ensuremath{{h}^{-1}\ {\rm kpc}}}

\newcommand{\omeganought}{\ensuremath{\Omega_{0}}}
\newcommand{\lambdanought}{\ensuremath{\lambda_{0}}}
\newcommand{\omegamnought}{\ensuremath{\Omega_{m0}}}
\newcommand{\omegalnought}{\ensuremath{\Omega_{\lambda0}}}

\begin{document}

\title{Early-Type Halo Masses from Galaxy-Galaxy Lensing}

\author{Gillian Wilson}

\address{Physics Department, Brown University, 182 Hope Street, Providence, RI 02912\\ 
E-mail: gillian@het.brown.edu}

\author{Nick Kaiser, Gerard A. Luppino and Lennox L. Cowie}

\address{Institute for Astronomy, University of Hawaii, 2680 Woodlawn Drive, Honolulu, HI 96822}  


\maketitle

\abstracts{We present measurements of the extended dark halo profiles 
of bright early-type galaxies at redshifts
$0.1 < z < 0.9$  
obtained 
via galaxy-galaxy lensing analysis of
images taken at the CFHT using the UH8K CCD mosaic camera.
Six $0.5 \times 0.5$ degree
fields were observed for a total of 2 hours each in $I$ and $V$,
resulting in catalogs containing $\sim 20 000$ galaxies per field.
We used $V-I$ color and $I$ magnitude to select bright early-type galaxies as the
lens galaxies, yielding a sample of massive lenses with fairly
well determined redshifts
and absolute magnitudes $M \sim M_* \pm 1$.
We paired these with faint galaxies lying at angular distances
$20'' < \theta < 60''$, corresponding to
physical radii of $ 26 < r < 77 \hkpc$ ($z = 0.1$) and $105 < r <315 \hkpc$ ($z =
0.9$), and computed the mean tangential shear $\gamma_{T}(\theta)$ of 
the faint galaxies.  
The shear falls off with radius roughly as $\gamma_{T} \propto 1/\theta$ 
as expected for flat rotation curve halos. 
The shear values were weighted in proportion to the square root
of the luminosity of the lens galaxy.  Our results
give a value for the average mean rotation velocity of an $L_\star$ galaxy 
halo at $r \sim 50-200\hkpc$ of
$v_\star =  238^{+27}_{-30}\kms$ for a flat lambda 
($\Omega_{{\rm m}0} = 0.3, \Omega_{\lambda 0} = 0.7$) cosmology
($v_\star = 269^{+34}_{-39}\kms$ for Einstein-de Sitter), and with 
little evidence for evolution with redshift.
We find a mass-to-light ratio of $\mlb \simeq 121\pm28h(r/100\hkpc)$ (for $\Lstar$ 
galaxies) and these halos
constitute $\Omega \simeq 0.04 \pm 0.01(r/100\hkpc)$ of closure density.
}

\section{Introduction}
\label{sec:intro}

Galaxy-galaxy lensing (the distortion of shapes of typically faint
background galaxies seen near typically brighter foreground galaxies)
offers a clean probe of the dark matter halo around galaxies. 
Here we shall restrict attention
to smaller scales where it is reasonable to interpret the results
as probing relatively stable and virialized halos of individual
galaxies.
Clusters of galaxies have traditionally been the primary target of weak
lensing studies. Individual galaxy masses are far more difficult to measure
due to their being less massive and hence yielding a smaller lensing
signal relative to the noise. 
However, by stacking pairs of galaxies it 
is possible to 
beat down the noise and measure the total average halo 
mass (characterized here by rotation velocity).


\section{THE DATA AND GALAXY SAMPLES}
\label{sec:data}

\subsection{Data Acquisition and Reduction}

The data were taken at the 3.6m CFHT telescope using the $8192 \times 8192$
pixel UH8K camera at prime focus. The field of view of this camera is 
$\sim 30 '$ with pixelsize $0.207''$. Six pointings were acquired as part
of an ongoing 
project whose principle aim is to investigate the cosmic shear pattern caused
by gravitational lensing from the large-scale structure  of the Universe.
This article is 
based on the second in a series of papers describing 
results from that project and focuses on properties of massive galaxy halos at radii of $20'' < \theta < 60''$ or $50 -200\hkpc$ (Wilson, Kaiser, Luppino \& Cowie~\cite{wklc-01} [Paper II]).
Kaiser, Wilson and Luppino~\cite{kwl-01} [Paper I] presented estimates of cosmic shear variance 
on $2' - 30'$ scales,  and Wilson, Kaiser \& Luppino~\cite{wkl-01} [Paper III] investigated
the distribution of mass and light on galaxy group and cluster scales. 


\subsection{Lens and Source Galaxy Samples}
\label{ssec:samples}

Our analysis differed from other groups in that we used $V -I$ color
to select a sample of bright early-type lens galaxies with reasonably
well determined redshifts.  As shown in \S~2.2 of Paper II, with
fluxes in 2 passbands and 
a judicious cut in red flux, one can reliably select bright early
type galaxies and assign them approximate redshifts.

To investigate the evolution of halo mass
with redshift, the lenses were firstly subdivided into
three slices of 
 width $dz = 0.3$ centered on redshifts 0.2, 
0.5 and 0.8 (Table~\ref{tab:results}). Secondly, a wider slice
of width $dz = 0.5$ centered on redshift 0.5 was analyzed.

\section{GALAXY DARK MATTER HALO MASSES}
\label{sec:results}

\subsection{Observed Tangential Shear Signal}
\label{ssec:gammatobs}

For each lens,
the mean tangential shear of faint `source' galaxies 
averaged over lens-source pairs
binned by angular separation is given by
\begin{equation}
\label{eq:gammaTdef}
\gamma_{T}(\theta) = - {
\sum\limits_{\rm pairs} W_l W_s M_{\alpha ij} \theta_i \theta_j \hat{\gamma}_\alpha / \theta^2
\over \sum\limits_{\rm pairs}  W_l W_s}
\end{equation}
where $\hat{\gamma}_\alpha$, for $\alpha = 1,2 $, is the shear estimate for the source galaxy,
$\theta$ is the projected 
angular separation of the lens and source,
$W_l$, $W_s$ are weights for the lens and source,
and the two constant
matrices $M_1$, $M_2$ are 
\begin{equation}
\label{eq:matrices}
M_{1lm} \equiv \left[ 
\begin{array}{cc}
1 & 0\\
0 & -1\\
\end{array}\right],
\;
M_{2lm} \equiv \left[ 
\begin{array}{cc}
0 & 1\\
1 & 0\\
\end{array}\right].
\end{equation} 

However, not all lens galaxies will contribute
equally to the shear signal. To optimize 
the signal to noise,
the shear contribution from each lens-source pair should be weighted 
by the mass of the lens. We assume here that
the Faber-Jackson relation ($M \propto \sqrt{L}$)
continues to larger radii and weight each lens accordingly. 
At each redshift, the resultant mean tangential shear signal 
falls off with radius roughly as 
$\gamma_{T} \propto 1/\theta$,
as expected for flat rotation curve halos. 

\subsection{Inferred Rotation Velocity}
\label{ssec:vstar}

For a flat rotation curve object the shear is given by
\begin{equation}
\label{eq:gammat}
\gamma_{T}(\theta) = \pi (v/c)^2 \langle \beta(z_l) \rangle / \theta 
\end{equation}
This equation allows one to convert 
between measured shear values and an equivalent rotation velocity
(the dimensionless quantity $\langle \beta(z_{l}) \rangle$ is calculated,
assuming a source galaxy redshift distribution based on spectroscopic
data from Len Cowie's ongoing Hawaii Deep Fields Survey).

As a result of the magnitude cut
discussed in \S~\ref{ssec:samples}
the inferred rotation velocity is for some effective luminosity $L_{\rm eff}$
galaxy. The equivalent mean rotation velocity, $\vstar$, for
an $\Lstar$ lens galaxy is computed using $\vstar^{2} = v^{2}/ \rtLeffrat$.

Column 2 of Table~\ref{tab:results} shows 
$\vstar$ at each redshift for a flat lambda ($\Omega_{{\rm m}0} = 0.3, \Omega_{\lambda 0} = 0.7$) cosmology.
We obtained values of $\vstar = 255^{+36}_{-42}\kms$ for $z = 0.2\pm0.15$,
$253^{+30}_{-35}$ for $z = 0.5\pm0.15$, and  $228^{+53}_{-70}$ for $z = 0.8\pm0.15$.
Thus, it appears that there is little evolution in the mass of dark matter halos 
with redshift. We then binned the signal for lens galaxies 
between $z = 0.25$ and $z = 0.75$ and concluded a rotation velocity
of
$\vstar = 238^{+27}_{-30}$ for $z = 0.5\pm0.25$. 

For comparison, column 3 again shows  
$\vstar$ but for an Einstein-de Sitter cosmology. The
inferred rotation velocity increases to $\vstar = 275^{+42}_{-50}$ 
,
$285^{+38}_{-44}$ and  $278^{+65}_{-85}$ for the same three intervals.
The increase in $\vstar$ in such a universe is primarily caused by smaller $\langle \beta \rangle$ values. 
We would conclude an overall rotation velocity of
$\vstar = 269^{+34}_{-39}\kms$ for $z = 0.5\pm0.25$ in this 
cosmology. 

\begin{table}[t]
\caption{Rotation velocity, $\vstar$, of an $L_\star$ galaxy as a function of redshift
and cosmology.\label{tab:results}}
\begin{center}
\footnotesize
\begin{tabular}{|c|cc|}
\hline
 & $\omegamnought = 0.3$ & $\omegamnought = 1.0$ \\
 & $\omegalnought = 0.7$ & $\omegalnought = 0.0$ \\
Lens Redshift  & \multicolumn{2}{|c|}{$\vstar$} \\
\hline
$0.2\pm0.15$  &  $255^{+36}_{-42}$  & $275^{+42}_{-50}$ 
\\
$0.5\pm0.15$  &  $253^{+30}_{-35}$  & $285^{+38}_{-44}$
\\
$0.8\pm0.15$  &  $228^{+53}_{-70}$  & $278^{+65}_{-85}$ 
\\
& & 
\\
$0.5\pm0.25$  &  $238^{+27}_{-30}$  & $269^{+34}_{-39}$ 
\\
\hline

\end{tabular}
\end{center}
\end{table}

\subsection{$M/L$ and Contribution to $\Omega_0$}

An $\Lstar$ galaxy halo with $\vstar = 238$ contains 
$1.31\times10^{12}(r/100\hkpc)h^{-1}\Mo$ within a 
radius of $r$ (since $M(r) = \vstarsq r/G$). An $\Lstar$ galaxy has a luminosity of 
$1.09\times10^{10}h^{-2}\Lo^{B}$, so
the mass to light ratio  is $\mlb = 121\pm28h(r/100\hkpc)$, or
about $\mlb \sim 250 h$ 
at the outermost points we can reliably measure.

The contribution of these early-type halos to the total density
of the Universe (again assuming that $M \propto \sqrt{L}$,
so $M(r) = M_\star(r) \sqrt{L / L_\star}$) is then
$\rho = M_\star(r) \int dL \; \phi_E(L) \sqrt{L / L_\star} = M_\star(r) \phi_{E\star} \Gamma(\alpha + 3/2)$.
This constitutes $\Omega = 0.04\pm0.01(r/100\hkpc)$ of closure density.

\section{Conclusions}

We used colors and magnitudes to cleanly select bright early-type galaxies.
By measuring a weighted mean tangential shear which 
decreased roughly as $1/\theta$ we concluded that early-type galaxies have 
approximately flat rotation curve halos extending out to several hundred $\hkpc$. By assuming a M $\propto \sqrt{L}$ relationship we inferred a rotation velocity
for an $\Lstar$ galaxy of 
$v_\star =  238^{+27}_{-30}\kms$ for $\omeganought = 0.3, \lambdanought = 0.7$
($v_\star = 269^{+34}_{-39}\kms$ for Einstein-de Sitter).
We sub-divided the galaxies and found little evidence for evolution with redshift. Finally, we determined a mass-to-light ratio for early-type halos of 
$\mlb = 121\pm28h(r/100\hkpc)$ (for $\Lstar$ galaxies) and found 
that these halos constitute 
$\Omega \simeq 0.04 \pm 0.01(r/100\hkpc)$ of closure density.

\end{document}